\documentclass[conference]{IEEEtran}
\usepackage{subcaption}
\usepackage{algpseudocode}
\usepackage{algorithm}
\usepackage[font=small]{caption}
\usepackage[pdftex]{graphicx}

\begin{document}
\title{A Survey of Bandwidth and Latency Enhancement Approaches for Mobile Cloud Game Multicasting}

\author{\IEEEauthorblockN{Mohammad Hosseini}
\IEEEauthorblockA{School of Computing Science\\
Simon Fraser University\\
Email: smhossei@sfu.ca}
}



\maketitle
\begin{abstract}
Among mobile cloud applications, mobile cloud gaming has gained a significant popularity in the recent years. In mobile cloud games, textures, game objects, and game events are typically streamed from a server to the mobile client.

One of the challenges in cloud mobile gaming is how to efficiently multicast gaming contents and updates in Massively Multi-player Online Games (MMOGs). This report surveys the state of art techniques introduced for game synchronization and multicasting mechanisms to decrease latency and bandwidth consumption, and discuss several schemes that have been proposed in this area that can be applied to any networked gaming context. From our point of view, gaming applications demand high interactivity. Therefore, concentrating on gaming applications will eventually cover a wide range of applications without violating the limited scope of this survey.
\end{abstract}
\IEEEpeerreviewmaketitle

\section{Introduction}
In the recent years, we see a tremendous growth in wireless networking enabled devices and this wireless capability and powerful processors have led to the mobile gaming industry  to  become  more  powerful,  capturing a progressively larger portion of  the  gaming  market. For instance based on the most recent statistics, it was predicted that the mobile gaming industry would reach \$18 billion by the end of 2014, including \%16.6 annual growth rate. It is estimated that 64 million people are playing mobile games monthly, which will rise to almost 95 million by the end of 2014 \cite{r23}. These mobile games mostly include multiplayer games. A multiplayer game is a game which is played by several players, mainly distributed players. These types of games heavily rely on wireless radios to communicate with other players. An example of a popular multiplayer game available for download to mobile devices is Quake.

There are some important issues related to mobile games. Multiplayer games need to send state updates, as a multicasting scheme to different players involved in the game. Actually every action an individual player performs typically needs to be communicated to all the related players and the players also need to react. So dissemination of each players actions and updates should be done in a scalable manner with very low latency. And on the other hand, due to the lack of processing power of the mobile devices, mobile games have some other limitations. Gaming uses a lot of power on these limited- resources mobile devices. Because of the game calculations and rendering, CPU is used more intensively. One of the most significant reasons for drains on the batteries of mobile devices is the wireless network interface card. It was shown that wireless network interface can account for up to \%70 of total power consuming in mobile devices. Thus addressing these issues is not profitable just from user tolerance perspective, but also beneficial in power saving and bandwidth saving aspects in the mobile devices which have constraints in resources.

In the following, we will present state-of-the-art approaches in cloud mobile gaming specifically targeting latency and bandwidth improvement, and will further discuss two of these approaches in more details.

\section{Proposed Approaches}
In this section, we present the state of the art ap- proaches used for efficiency improvements in networked mobile games, in order to make the gaming experience more efficient, gaining the most user tolerance and resource saving.

\subsection{Multiple Transmission}
Multiple packet  transmission  is  a  general  approach that could be applied to any networked game. It is a simple method used for transmitting the lost packet two or three times \cite{r24}. However, using this approach would certainly reduce the  mobile network bandwidth, but it could be rather effective for receiving some important and necessary information such as avatar state changes.

Thus it is required to employ some strategies for detecting important data and related packets.

\subsection{Avatar Blinking}
Avatar Blinking is a game-specific approach used for giving an uncertainty effect about the future of an affected avatar by blinking it. Consider a situation in which a local player shoots another avatar. In such cases, the player has to wait for the shooting outcome from the other avatar. Using Avatar Blinking gives the player a positive and rapid feedback which could be effective rather than being unresponsive. Eventually, the response for the effects would arrive after a few seconds and then the avatar could leave behind the blinking status \cite{r25}.

\subsection{Dead-Reckoning}
When packets containing some useful game information have failed to arrive at the client, we try to guess the avatars movement or rotation. The approach is called Dead Reckoning. We choose to activate DR after one movement packet is lost, and to keep it switched on for at most ten screen updates.

Dead Reckoning (DR) requires the receiving packets to be time-stamped. It is also essential for the client to have an estimation of how long to wait before a packet is deemed to be lost. We must also provide some strategies for dealing with a lost packet turning up after a lengthy latency. In case of having a packet loss, Dead Reckoning would start on; so instead of unresponsiveness, we see the avatar will still keep moving. However, the movement predictions may be very hard. So using Dead Reckoning will be switched off after some updates \cite{r26, r27}.

\subsection{Blue Banana}
Blue Banana is an approach for modeling and anticipation of avatars mobility. If we do not anticipate this mobility, the needed information and thus, the corresponding packets are not delivered on time, which leads to transient failures at the client-side application level. We see current Multiplayer Online Games based on peer-to-peer overlays render the virtual world surrounding the player avatar very slowly. This happens because avatar movement implies many data exchanges through the overlay. Blue Banana employ a mechanism to predict avatar movement which based on the results, permits to load 20 times more data about playing  areas  in case of mobility. The algorithm decreases the number of transient failures  of  a  moving  avatar:  the  loading of  the  data  composing  the  forthcoming  playing  areas
 
begins earlier which permits the construction of a correct playing area image when the avatar effectively enters inside it \cite{r28}. Utilization of Blue Banana as a modeling scheme for massively multiplayer mobile games would lead to less response time and efficiency improvements.

\subsection{COPSS-based Gaming}
Gaming over a Content- Oriented Publish/Subscribe system or COPSS-based Gaming in brief, is a very novel client- server approach centered on content distribution instead of host-to-host connectivity. This change from host centric to content centric separates the action receivers from the performers. It is applied on massive multiplayer games as a push-based multicasting (recall, we have two approaches for multicasting, push-based and pull- based), wherein the players can directly push the content to the other players in the same sub-world in an efficient manner. The role of the server is greatly reduced in COPSS-based Gaming \cite{r29}].

In the following parts, we will discuss two of the approaches explained above, Dead-Reckoning as the mostly used approach, and also COPSS-based Gaming as the newest approach.

\section{Detailed Discussion}

\subsection{Dead-Reckoning}
As we mentioned before, Dead Reckoning (DR) is actually a latency hiding approach used in networked games mostly in multiplayer games. In DR we try to guess and predict behavior and the state of game objects in case of packet failures.

\begin{figure}[!t]
\caption{An example of Dead Reckoning mechanism}
\centering
\includegraphics[width=1\columnwidth]{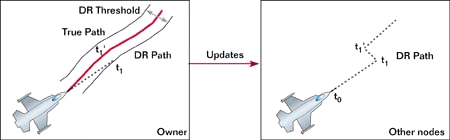}
\label{dr}
\end{figure}

DR is used as a technique for both latency hiding and bandwidth reduction. It is a form of replicated computing in which every player involved in a game simulates all the other players entities, like the enemies of first players avatar.

The basic notion of dead reckoning is that before starting to play, the players should agree on a set of algorithms that can be used by all player nodes to extrapolate the behavior of entities in the game, and also they should come into an agreement on how far reality should be allowed to get from these extrapolation algorithms before a correction is issued. The deviation between actual and extrapolated trajectories is known as the dead reckoning error.

Figure \ref{dr} shows an example on how to use dead reckoning. As shown in Figure \ref{dr}, for each playing avatar, we have to maintain two parallel models, an extrapolated path (which is often called ghost model) and also an actual path that is the true state as computed from player inputs. The ghost model represents an approximation of the true entity state. Game state updates are used to ensure consistency between the two models. The local client compare the dead reckoning error to a defined error threshold (which is a part of in advance agreement we mentioned before) to dynamically determine whether or not the game object deviates too far from the real trajectory and thus requires a game-state update. Another note is that as mentioned in the suggested approaches part, DR mechanisms often use convergence techniques to smoothly correct the current state which was caused from inaccurate extrapolation. This smooth state changing is so significant in the point of user gaming experience, so not to suddenly jump from an inaccurate point to the true point \cite{r30}.

Besides using DR for latency hiding and bandwidth saving, we can even use it as an option for power saving. Lowering the power used by the wireless interface would greatly decrease the overall power consumed by the mobile device. This is achieved by spending as much time as possible  in  low-power  states.  A  recent  work in Network Systems Lab at Simon  Fraser  University has shown that we can use dead reckoning error rates to dynamically devise a power saving schedule for the wireless interface to predict periods of inactivity in the wireless device till the next  update  to  happen,  so  to put it to a sleep mode. The results showed we could save up to \%36 of mobile battery power if we use DR for power saving \cite{r27}. So DR can be used not just for latency and bandwidth improvement, we can even use dead reckoning to save power, which is so profitable in mobile devices, having limited resources.

\subsection{COPSS-based Gaming}

\begin{figure}[!t]
\caption{(I) An example for a hierarchical game. (II) Assigning IDs to different sub-worlds.}
\centering
\includegraphics[width=1\columnwidth]{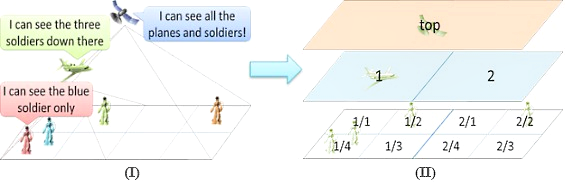}
\label{cops}
\end{figure}

As we described in the previous section, COPSS is the short form for Content-Oriented Publish/Subscribe System. In such systems, we leverage a push-based multicasting scheme with COPSS-aware routers as in the infrastructure which relies on publishing and subscribing to a multicasting group. COPSS-based gaming is used mostly on Multiplayer online roleplaying games, in which the gaming needs high interactivity and persistent view of the environment, which is called world in a gaming perspective. World of Warcraft, Counter- Strike and even Second Life (However Second Life is not actually a gaming environment) are examples of such games which are actually characterized by the need for a short latency since every action performed by each player needs to be multicasted to all other players viewing the same sub-world.

The basic assumption we make in this approach is that all the players share a same map. For practical reasons such as efficient broadcast of updates, the world map is divided into various zones and each of these zones can be divided further into areas and smaller areas, thereby forming a hierarchy. In this hierarchical map, different players have different amount of visibility; this is what we see in massive multiplayer online role-playing games, called MMORPG. Figure \ref{cops} (I) shows an example  for such game in which different players depending on their sub-world, have different amount of visibility, and as Figure \ref{cops} (II) shows we assign a unique ID to each sub- world.

\begin{figure}[!t]
\caption{Logical hierarchy, transformation of the gaming system into a tree.}
\centering
\includegraphics[width=1\columnwidth]{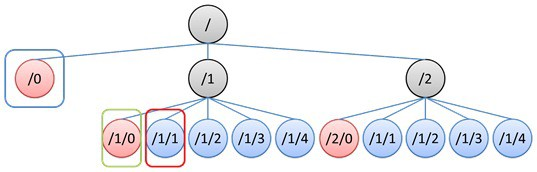}
\label{cops2}
\end{figure}

In  order  to  implement  a  COPSS-based  system  in  a game, we should consider the gaming environment and the above mentioned map hierarchy as a tree, along with publishing and subscribing actions. So as the protocol, based on the visibility and update needs of each player, for publishing, each node (player) would publish to the subtree of the area the he is in, and for subscribing, in order to receive game updates, each node (player) would subscribe both to the area he is in, and all the /0s along the hierarchy to the upper layers. a /0 node is the representative for the upper level. Figure \ref{cops2} shows the corresponding paradigm.

Eventually, if we implement this numbering method in our example gaming system, as mentioned in Figure \ref{cops}, we would have a location ID for each player, and a publishing/subscribing ID in order to send to and receive game updates from the appropriate multicasting group.

\begin{figure}[!t]
\caption{Location ID and subscription/publishing IDs of players after employing COPSS for a game.}
\centering
\includegraphics[width=1\columnwidth]{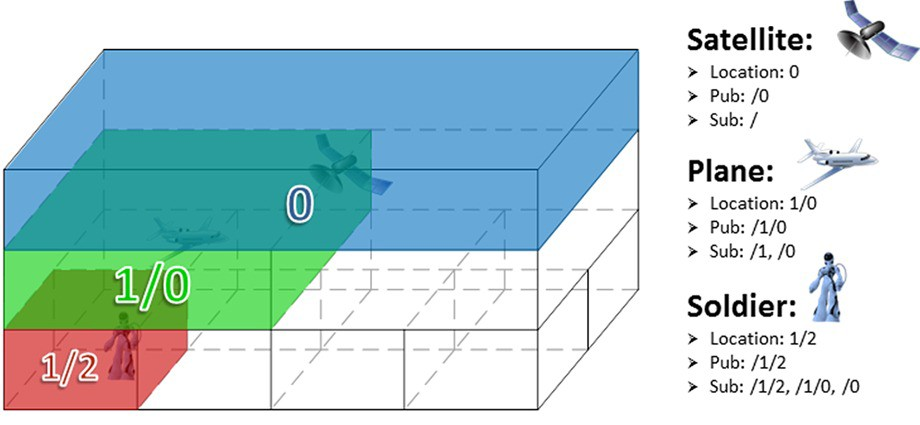}
\label{cops3}
\end{figure}

Figure \ref{cops3} shows IDs for a selected number of players. Table I shows the performance of COPSS-based gaming based on a trace-driven simulation. This trace is taken from the transmitted packets in a busy Counter-Strike game server \cite{r29}. According to this table, we see G- COPSS disseminates updates within 1/10 of a single s, while in the same condition, the IP server-based solution as the state-of-the-art solution takes more than 21s and uses up to almost 1.8 times more network traffic due to server congestion.

\begin{table}[!t]
\caption{Performance comparison of G-COPSS and IP-Server}
\centering
\begin{tabular}{|c|c|c|}
\hline 
 & Update Latency (ms) & Network Load \\ 
\hline 
G-COPSS & 94.885 & 5,557,641 \\ 
\hline 
IP Server & 21,448.168 & 9,622,658 \\ 
\hline 
\end{tabular} 
\end{table}

\section{Conclusion}
In the recent years, there has been a growing trend towards cloud gaming services to mobile devices. Because of the resource limitations of the mobile devices, employment of efficient approaches for game multicast services is crucial.

In this survey, we discussed the state of the art approaches for cloud game content multicasting over mobile networks. We first introduced a wide category of approaches in the domain, and further discussed two major and novel approaches used for multicasting game updates on mobile devices.



\bibliographystyle{IEEEtran}
\bibliography{refs}

\pdfinfo{
   /Author (Mohammad Hosseini et al.)
   /Title  (ICC Paper)
   /Subject (PDFLaTeX)
   /Producer (PDFLaTeX)
   /Keywords (PDF;LaTeX)
}
\end{document}